# A new analytical model for the contact of Gaussian rough surfaces

Sihe Wang[1], Weike Yuan[1], Xuanming Liang, Gangfeng Wang[*]

Department of Engineering Mechanics, SVL and MMML, Xi'an Jiaotong University, Xi'an 710049, China

[1] These authors have equal contribution.

[*] Corresponding author. E-mail: wanggf@mail.xjtu.edu.cn

**Abstract:** This paper advances an analytical incremental contact model for the purely elastic or elastic-perfectly plastic Gaussian rough surfaces. The contact is modelled by the accumulation of identical circular contacts with radius given by the total truncated area at varying heights divided by the contact patch number. The contact area-load relationship is derived analytically, showing approximate linearity for the contact fraction up to 10%. Good agreement is found between the new proposed model and the direct finite element simulations. To characterize the influence of plastic deformation, a dimensionless plasticity parameter is introduced as the ratio of yield strain to root mean square gradient of the rough surface. It is demonstrated that the general elastic-plastic contact response would approach to the limit of purely elastic as the plasticity parameter increases.

# 1. Introduction

Nearly all of practical solid surfaces are rough spanning over multiple scales [1,2]. Owing to the nature of surface roughness, mechanical contacts of solids usually take place partially at a series of randomly distributed, different sized and irregularly shaped spots. The real contact area is generally a small fraction of the nominal area. For its essential role in many physical phenomena, like friction, wear, sealing, thermal and electrical conductance, contact mechanics of rough surfaces has long been recognized as a critical fundamental problem in tribology [3,4]. In particular, an accurate prediction of the total real contact area at given external load is of great importance.

Over the last decades, numerous theoretical approaches have been developed to deal with this issue, including various multi-asperity contact models [5-8] and the famous Persson's theory [9]. The basic idea of multi-asperity model originated from the pioneering works by Archard [5] and Greenwood and Williamson (GW) [6]. The GW model [6] assumed that the rough surface can be modelled as an ensemble of identical spherical asperities with a Gaussian distribution of height, and each asperity coming in contact independently can be described by Hertz theory [10]. Later, Bush et al (BGT) [7] suggested that the asperities should be treated as paraboloids. The statistical distribution of asperity height and curvatures was given by a random process model for isotropic Gaussian rough surfaces [11]. Because of the non-axisymmetry of asperity, their solution takes a rather complicated form involving full calculation of triple integrals. Instead, Greenwood [8] adopted an approximate solution for elliptical Hertzian contacts based on the geometric mean curvatures of asperities and reproduced the BGT solution in a simplified manner. Despite of the popularity, it has been aware there are significant drawbacks in these multi-asperity models. For example, the interaction and coalescence between neighboring asperities were not taken into account so that these models were considered to be valid only for ultimately small contact area. For this reason, a lot of effort has been put into the modification of multi-asperity models [12-15], most of which require iterative numerical computations on the contact behavior of every single asperity. In another completely different way, Persson [9] developed a general theory for the contact of solids with multi-scale surface roughness, in which the surface power spectral density is required, irrespective of the specific geometry of asperities. It was derived that the dependence of probability distribution of contact pressure on the magnification satisfies a diffusion type equation at the full contact state. Partial contact was solved by imposing a boundary condition that the probability distribution vanishes at zero-pressure for non-contact parts. Also, Persson's theory has its weakness, especially that the prediction becomes less accurate as the contact is far from the full contact state [16]. Carbone and Bottiglione [17] compared various multi-asperity models and Persson's theory. Interestingly, both two kinds of model predict a linear relation between the real contact area and the nominal pressure in the limit of large surface separation, but in different proportionality.

Considering the fact that the real area in intimate contact is small and discrete, the local effective stress at or near contact spots can be far beyond the material elastic limit, even at a very light load. Both numerical and experimental studies showed that plastic deformation is inevitable throughout the contact process of rough surfaces [18,19]. To account for the plasticity, the concepts of multi-asperity model were widely employed to model the elastic-plastic contact behavior [20-23]. For example, Chang et al [20] extended the original GW model [6] to the elastic-plastic case by considering the volume conservation of plastically deformed asperities and applying Hertz theory for the elastically deformed asperities. For the rough surfaces characterized by fractal function, Majumdar and Bhushan [21] applied Hertz theory or the fully plastic contact model to each contacting asperity, according to the corresponding asperity contact area is larger or smaller than a critical value. They assumed that a power-law relationship, which is originally derived for the size distribution of islands on the earth surface [24] (analogously, for the geometrically truncated areas of asperities on the rough surface), can be utilized to describe the size distribution of actual contact patches. In addition, Persson [25] extended his theory to the elastoplastic contact case using the yield stress as the upper bound of local contact pressure. These elastic-plastic contact models still display a linear rise of contact area with the load increasing, though appreciable plastic deformation is considered

Moreover, the limit of fully plastic contact is worthy of note. Bowden and Tabor [26] first suggested to model the contact of rough surfaces with completely plastic flow, in which a direct linear load-area relationship was obtained for explaining the classical friction law. Pullen and Williamson [27] noticed the significant effect of asperity interaction and proposed an energy balance method to study the fully plastic contact of rough surfaces with volume conservation. They found that the linearity between real contact area and load breaks down at high loads, which is consistent with their experiments. Based on the observations of Pullen and Williamson [27], Nayak [28] further attempted to develop a random process model of Gaussian rough surfaces in plastic contact. Probably because the strong assumption of fully plastic deformation was generally unrealistic, these works did not attract much attention. However, they still provide important information for the understanding of rough surface contact mechanics. As Greenwood stated [29], Nayak's work [28] *does tackle a fundamental question of contact between rough surfaces*. Specifically, it was suggested that the number of contact patches and the total real contact area can be analyzed based on the intersection of rough surface and a virtual plane. Referred to as the profilometric theory [30], this idea was also introduced partly in the preceding work of Abbott and Firestone [31], which was originally aimed to describe the wear process of rough surfaces. Recently, using the profilometric theory to determine the real contact area and further to evaluate the contact response of rough surfaces was greatly appreciated [32-34]. It was demonstrated that extension of the profilometric theory beyond fully plastic contact can be achieved successfully by using an incremental equivalent approach [33,34]. This method treats the contact of rough surfaces as an accumulation of identical circular contacts with radius estimated from the total contact area and the number of contact patches. As yet, this model is still a deterministic one, which demands substantial numerical analysis on the geometry of intersections of a given rough surface at different heights.

Herein, based on the random process model of Nayak [28] and the profilometric theory [30], an analytical version of the incremental equivalent circular contact model is formulated for the contact between Gaussian rough surfaces and rigid flats. This new proposed model is applicable to both the purely elastic contact and the elastic-perfectly plastic contact.

## 2. Analytical contact model

Fig. 1(a) shows the problem under investigation that a deformable solid with randomly rough surface is compressed by an ideally rigid flat plane. A Cartesian coordinate system (*O-xyz*) is constructed to fix positions of the surface points, where *x*-axis and *y*-axis are taken in the mean plane of the rough surface ($z = 0$). Assume that the rough surface can be considered as an isotropic Gaussian random process with power spectral density of $C(q)$. Then, the zeroth, second and fourth spectral moments of the random process can be defined as

$$m_0 = 2\pi \int_{q_0}^{q_1} C(q) q \,\mathrm{d}q, \quad m_2 = \pi \int_{q_0}^{q_1} C(q) q^3 \,\mathrm{d}q, \quad m_4 = \frac{3\pi}{4} \int_{q_0}^{q_1} C(q) q^5 \,\mathrm{d}q \tag{1}$$

respectively, where the root of $m_0$ equals the root mean square surface roughness $\sigma_0$, and the root of $m_2$ equals the root mean square gradient. To characterize the spectral breadth of random rough surfaces, a dimensionless combination as $\alpha = m_0 m_4 / m_2^2$ is usually introduced [11].

For a Gaussian rough surface, the probability density function for the surface height is given by

$$p(z) = \frac{1}{\sqrt{2\pi m_0}} \exp\left(-\frac{z^2}{2m_0}\right) \tag{2}$$

Consequently, the truncated area of the rough surface at a specified height $z$ can be obtained by [30]

$$\frac{A(z)}{A_0} = \int_z^\infty p(z)\,\mathrm{d}z = \frac{1}{2}\mathrm{erfc}\left(\frac{z}{\sqrt{2m_0}}\right) \tag{3}$$

where $A_0$ is the nominal area, and erfc($x$) is the complementary error function defined by

$$\mathrm{erfc}(x) = \frac{2}{\sqrt{\pi}} \int_x^\infty e^{-y^2}\,\mathrm{d}y \tag{4}$$

According to Nayak's random process model [28], a lot of closed contours in $A(z)$ are indeed not singly connected; namely, there are "holes" in the contact patches. Nonetheless, the density of such holes becomes negligible for large surface separation or small contact fraction, which is particularly apt to arrive for narrow-band rough surface with small $\alpha$ [28,29]. Neglecting the density of holes in the first approximation, the density of contact patches can be expressed as [28,29]

$$\frac{N(z)}{A_0} = \frac{1}{(2\pi)^{3/2}} \frac{m_2}{m_0} \frac{z}{\sqrt{m_0}} \exp\left(-\frac{z^2}{2m_0}\right) \tag{5}$$

Recently, our group proposed an incremental approach to calculate the area-load relation for the contact of rough surfaces [33,34]. In this method, the truncated area is treated as the real contact area and the irregular contact patches are further simplified by a group of identical circular patches, as shown in Fig. 1 (b) and (c). And the equivalent contact radius is given by

$$R(z) = \sqrt{\frac{A(z)}{\pi N(z)}} \tag{6}$$

which will be used to determine the contact stiffness at a specified surface separation.

For a decrement of surface separation d$z$, the corresponding increment of load d$P$ can be approximately expressed as

$$dP = N(z)k(z)dz \qquad (7)$$

where $k(z)$ is the contact stiffness of a circular contact with radius $R(z)$.

Subsequently, the load for generating a certain real contact area can be obtained by an accumulation process with the height decreasing from infinity to $z$. In this process, both the area $A(z)$ and the number of contact patches $N(z)$ at different height are required. Instead of using complete numerical technique as in [33,34], this work would apply analytical forms for the total area and number of contact patches. In what follows, the purely elastic contact and elastic-perfectly plastic contact of Gaussian rough surfaces will be analyzed separately with the help of this incremental equivalent approach.

*(a) Purely elastic contact*

First, we suppose there only exists elastic deformation during the compression of a rough surface with Young's modulus $E$ and Poisson's ratio $v$. In this case, the contact stiffness of each equivalent circular contact patch is given by $k(z) = 2E^*R(z)$, where $E^* = E/(1-v^2)$ [35]. Substituting this contact stiffness into Eq. (7) and combing Eqs. (3)-(7), one can obtain the load satisfying the following equation,

$$\frac{d\overline{P}}{dt} = \Phi(t) \qquad (8)$$

where $\overline{P} = P/(A_0 E^* \sqrt{m_2})$, $t = z/\sqrt{m_0}$, and $\Phi(t)$ is a dimensionless function given by

$$\Phi(t) = \frac{1}{(2\pi^5)^{1/4}} \sqrt{t \, \mathrm{erfc}\left(\frac{t}{\sqrt{2}}\right) \exp\left(-\frac{t^2}{2}\right)} \qquad (9)$$

For a compression with the height decreasing from infinity to $z$, the load $\overline{P}$ is obtained by integration of Eq. (8),

$$\overline{P} = \int_t^\infty \Phi(t) dt \qquad (10)$$

From Eq. (3) and Eq. (10), the relationship between load and real contact area can be established by eliminating the intermedium variable $t$. Asymptotically, the load is proportional to the contact area as $P \to 0$,

$$\frac{P}{A_0 E^* \sqrt{m_2}} = \frac{\sqrt{2}}{\pi} \frac{A_c}{A_0} \tag{11}$$

*(b) Elastic-perfectly plastic contact*

For the elastic-plastic contact of rough surfaces, the determination of contact stiffness of the circular contact patches for each incremental step is not straightforward. Here we consider an elastic-perfectly plastic rough solid with yield stress $\sigma_Y$. Based on the finite element simulations and parametric analysis, it was found that the contact stiffness can be approximately fitted by [34]

$$k(z) = 2E^* R(z) \cdot g\left(\frac{p_m}{\sigma_Y}\right) \tag{12}$$

where $p_m = P/[N(z) \cdot \pi R(z)^2]$ is the average contact pressure within the real contact area, and $g(x)$ is a dimensionless empirical function given by

$$g(x) = \left(1 + 0.0415 x^{4.50}\right)^{-1.40} \tag{13}$$

Similarly, using Eqs. (3)-(6) in Eq. (7) and substituting for the contact stiffness from Eq. (12), we obtain

$$\frac{d\bar{P}}{dt} = g(\Lambda) \Phi(t) \tag{14}$$

with

$$\Lambda = \frac{\bar{P}(t)}{\eta A(t)/A_0} \tag{15}$$

$$\eta = \frac{\sigma_Y / E^*}{\sqrt{m_2}} \tag{16}$$

A numerical algorithm based on Runge-Kutta method can be used to solve this differential equation. The mathematical initial condition of $\bar{P} = 0$ as $t \to \infty$ is replaced by the truncated initial condition of $\overline{P}(t_0) = 0$ with $t_0$ being a sufficiently large upper limit. Again, the load-contact area relation can be derived with the contact area given by Eq. (3) and the load determined by Eq. (14) at different height $t$. Interestingly, the asymptotic linearity between contact area and load also exists in this case. Denote the slope as $S = P/(A_c E^* \sqrt{m_2})$, and assume it keeps constant as $P \to 0$. Then, we have

$$S = \frac{\sqrt{2}}{\pi} g\left(\frac{S}{\eta}\right) \tag{17}$$

By solving the algebra equation numerically, one can determine the slope $S$ of the asymptotic load-area relation.

## 3. Results and discussions

To test this analytical model, the contact problem considered above is also examined by using the finite element method. Similar simulations as in [33,34] are performed for the Gaussian random rough surfaces, which are artificially generated by using an open-source code in MATLAB [36]. For convenience, the rough surfaces are set to be self-affine fractal and have the following power spectral density,

$$C(q) = \begin{cases} C_0 \left(\dfrac{q}{q_0}\right)^{-2(1+H)}, & \text{for } q_0 \leq q \leq q_1 \\ 0, & \text{elsewhere} \end{cases} \tag{18}$$

where $C_0$ is a constant, $H$ is the Hurst exponent, $q_0$ and $q_1$ are the lower and upper wavevector cutoff, respectively. Note that the constant $C_0$ can be obtained through Eq. (1) as $\sigma_0$, $H$, $q_0$ and $q_1$ are given.

With the input parameters listed in Table 1, three different rough surfaces (A, B, and C) are generated, of which the spectral breadth parameters $\alpha$ are 12.28, 65.02, 100.64, respectively. For the synthetic rough surfaces, Fig. 2 displays their topographies, height distributions, numerically obtained truncated contact area and density of contact patches at different height. It can be seen that the synthetic surfaces are of good Gaussianity and Eq. (3) can accurately characterize the actual truncated contact area at different height. In addition, the density of contact patches is in reasonably consistent with the analytical prediction of Eq. (5) for $t > 1.28$, which is corresponding to the area fraction $A(z)/A_0$ smaller than about 10%. With these conditions satisfied, our analytical model is ready to predict the contact behavior of the synthetic rough surfaces.

For the purely elastic contact of Gaussian rough surfaces, our analytical model predicts a one-to-one relationship between the contact area fraction $A_c/A_0$ and the normalized load $P/(A_0 E^* \sqrt{m_2})$. As shown in Fig. 3, the repeatedly demonstrated linearity between contact area and load approximately holds in the present model, even for a large contact fraction up to 10%. For comparison, Fig. 3 also plots the results of finite element simulations for three synthetic surfaces. Clearly, good agreement with our model can be found.

In this case, the predictions of BGT model [7] and Persson's theory [9] are worthy of mention, which were generally quoted to compare with the newly developed contact models. It should be noted that both Persson's theory and our analytical model are irrelevant to the spectral breadth parameter $\alpha$, while the fully calculated BGT model depends strongly on the value of $\alpha$. Thus, the asymptotic BGT solution [7] is presented for comparison. For a given load, the contact fraction predicted by our model lies above the predictions of BGT model and Persson's theory. In the limit of vanishing contact area, our model gives an asymptotic solution for the mean contact pressure as $P/(A_c E^* \sqrt{m_2}) = \sqrt{2}/\pi$. This is lower than $1/\sqrt{\pi}$, according to BGT model [7], and further lower than $\sqrt{\pi}/2$ given by Persson's theory [9].

For the contact of elastic-perfectly plastic Gaussian rough surfaces, it can be found from Eq. (14) that the contribution of plasticity is governed by the dimensionless function $g(\Lambda)$. For different values of the parameter $\eta$, Fig. 4 plots the variation of $g(\Lambda)$ with respect to the normalized load $\overline{P}$. It can be seen that $g(\Lambda)$ is basically insensitive to the variation of load, but varies remarkably with $\eta$. A larger $\eta$ results in a higher level of $g(\Lambda)$, implying that the portion of plastic contact deformation gets lower. For example, the value of $g(\Lambda)$ is greater than 0.98 when $\eta > 0.5$, which indicates that the contact is mainly elastic. Thus, if a rough surface is especially smooth, or has ultrahigh plastic yield strain, its overall contact response could be dominated by elasticity across the whole contact process. It is worth mentioning that Greenwood [8] had also introduced a similar parameter, named plasticity index $\psi_M$, where $\psi_M^{-1} \approx 2.7\eta$, to measure whether the contact is primarily elastic, and suggested that $\psi_M$ should be smaller than 0.5 for the primarily elastic contact, corresponding to $\eta > 0.75$.

The contact area-load relationship for the contact between a rigid flat and an elastic-perfectly plastic substrate with rough surface is displayed in Fig. 5 for $\eta = 0.04, 0.1, 0.2,$ and 0.5. Our model represented by thin lines again agrees well with direct finite element simulations for surface B. Even with plastic deformation, the real contact area is approximately proportional to the applied load, but the slope is certainly dependent on the plasticity parameter $\eta$. Such dependence can be reasonably described by the asymptotic solution of Eq. (17). As discussed above, the influence of material plasticity increases with the value of $\eta$ decreasing. To generate the same contact area in flatting the rough solid with a larger plasticity parameter $\eta$ requires a higher load. In other words, the rough surface with a smaller $\eta$ would have a lower mean contact pressure. This is intuitively reasonable considering the fact that a relatively flat rough surface with small $m_2$ would have lower level of stress concentration in the contact interface, and the fact that a relatively stiff substrate with large $\sigma_Y/E^*$ would be more difficult to occur plastic yield. As expected, the contact response of a rough surface with sufficiently large $\eta$ would ultimately return to the solution of purely elastic contact.

## 4. Conclusion

In summary, we present an analytical version of the incremental equivalent circular contact model to determine the area-load relationship for Gaussian rough surfaces. Both the purely elastic contact and the elastic-perfectly plastic contact are considered, and the predictions of the new model are in good agreement with direct finite element simulations. It is found that the influence of material plasticity can be evaluated by a parameter $\eta$ defined as the ratio of yield strain to root mean square gradient. Even for the contact up to 10% of the nominal area, the contact area is approximately proportional to the load, and the proportionality of general elastic plastic contact would approach to the limit of purely elastic contact as the value of $\eta$ increases. The contact of a rough surface can be considered to be mainly elastic if its $\eta$ is larger than 0.5. This work provides an effective candidate for the modelling of normal contact of Gaussian random rough surfaces.

**Acknowledgement**

Supports from the National Natural Science Foundation of China (Grant No. 11525209) and the SVL Program (No. SV2021-ZZ-18) are gratefully acknowledged.

**Table caption**

Table 1. Input parameters for generating Gaussian rough surfaces

**Figure captions**

Fig. 1 Schematics of the contact problem. (a) Compression of a Gaussian rough surface by a rigid plane. (b) Actual irregular contact patches. (c) Equivalent circular contact patches.

Fig. 2 Synthetic rough surfaces: surface topographies, height distributions, truncated contact area and density of contact patches, respectively.

Fig. 3 The contact fraction $A_c/A_0$ as a function of the normalized load $P/(A_0 E^* \sqrt{m_2})$ for elastic contact.

Fig. 4 Variation of $g(\Lambda)$ with load for $\eta = 0.04, 0.1, 0.2$, and $0.5$.

Fig. 5 The contact fraction $A_c/A_0$ as a function of the normalized load $P/(A_0 E^* \sqrt{m_2})$ for elastic- plastic contact.

Table 1

| Surface | $\sigma_0$ /μm | $H$ | $q_0$ /m$^{-1}$ | $q_1$ /m$^{-1}$ |
|---|---|---|---|---|
| A | 0.4 | 0.9 | 2.51e5 | 4.02e6 |
| B | 0.5 | 0.5 | 3.14e4 | 4.02e6 |
| C | 0.5 | 0.6 | 3.14e4 | 4.02e6 |

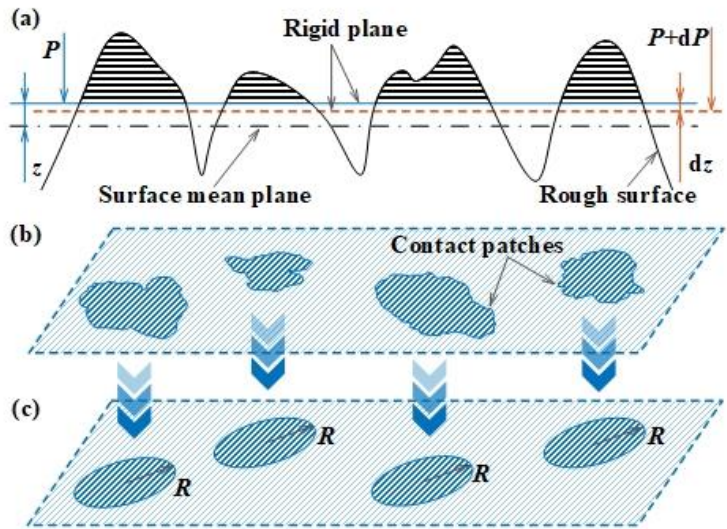

**Fig. 1**

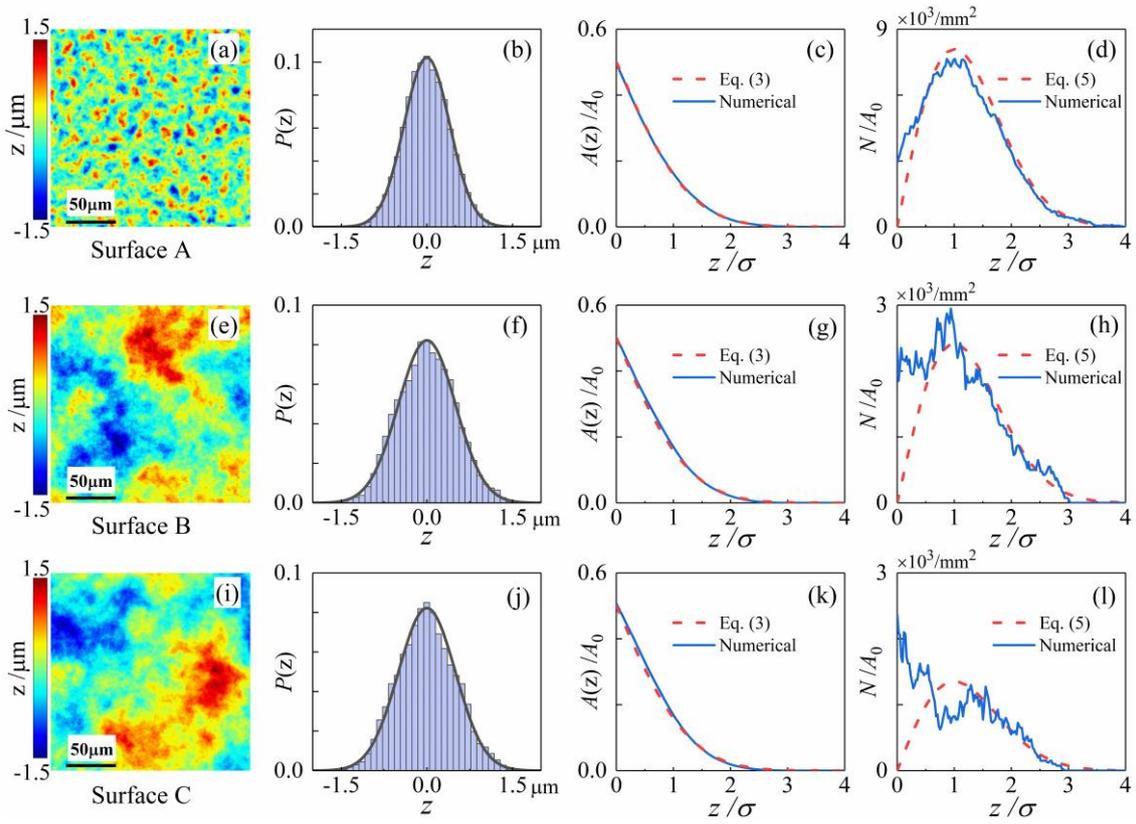

**Fig. 2**

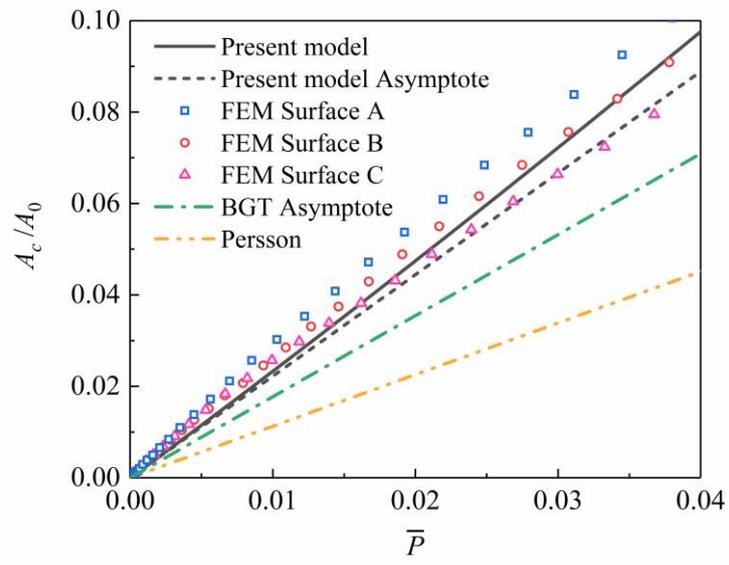

**Fig. 3**

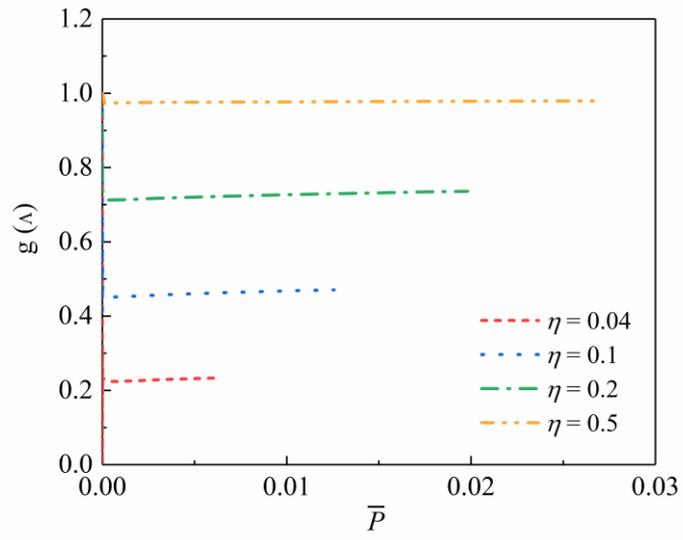

**Fig. 4**

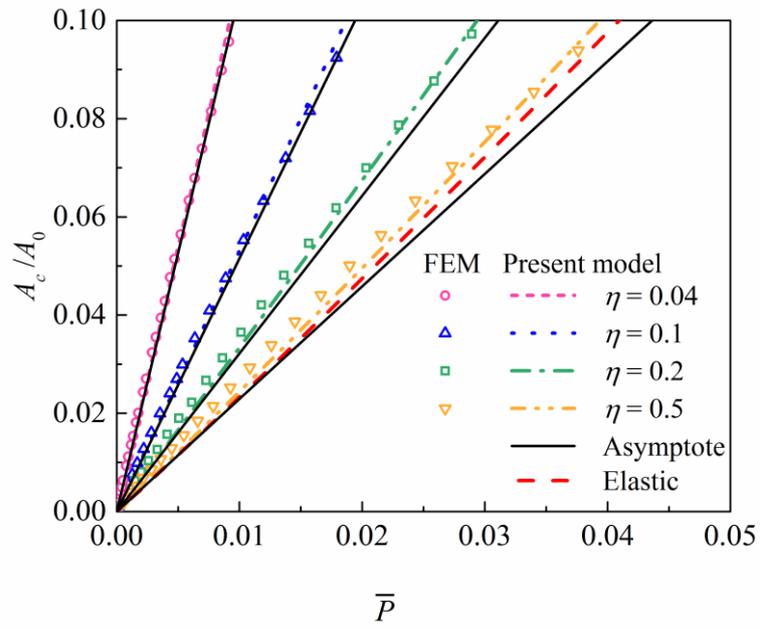

**Fig. 5**